\documentclass[12pt]{article}

\usepackage{newtxtext,newtxmath}

\usepackage{graphicx}

\usepackage[letterpaper,margin=1in]{geometry}

\renewenvironment{abstract}
	{\quotation}
	{\endquotation}

\date{}

\makeatletter
\renewcommand{\fnum@figure}{\textbf{Figure \thefigure}}
\renewcommand{\fnum@table}{\textbf{Table \thetable}}
\makeatother

\usepackage{scicite}

\usepackage{url}

\def\scititle{
	Single-exposure holographic 3D printing via inverse-designed phase masks
}
\title{\bfseries \boldmath \scititle}

\author{
	Dajun~Lin$^{1\dagger}$,
	Xiaofeng~Chen$^{2\dagger}$,
	Connor~O. Dea$^{2\dagger}$,
    Ji-Won~Kim$^{2}$,
    Keldy S.~Mason$^{2}$,\and
    Kwong Sang~Lee$^{2}$,
    Apratim~Majumder$^{1}$,
    Chih-Hao~Chang$^{3}$,
    Michael~Cullinan$^{3\ast}$,\and
    Zachariah A.~Page$^{2\ast}$,
    Rajesh~Menon$^{1\ast}$\and
	\small$^{1}$Department of Electrical \& Computer Engineering, University of Utah, Salt Lake City \& 84112, USA.\and
	\small$^{2}$Department of Chemistry, University of Texas at Austin, Austin \& 78712, USA.\and
    \small$^{3}$Walker Department of Mechanical Engineering, University of Texas at Austin, Austin \& 78712, USA.\and
	\small$^\ast$Corresponding authors: rmenon@eng.utah.edu, zpage@cm.utexas.edu, michael.cullinan@austin.utexas.edu\and
	\small$^\dagger$These authors contributed equally to this work.
}

\begin{document} 

\maketitle

\begin{abstract} \bfseries \boldmath
Additive manufacturing using lasers is typically constrained by serial, voxel-by-voxel or layer-by-layer processing, which fundamentally limits printing speed. Here, we introduce a single-shot holographic three-dimensional (3D) printing technique that overcomes this bottleneck through the combined use of inverse-designed microstructured phase masks and photopolymer resins with tailored optical absorption. By precisely engineering the topography of the phase mask, we sculpt the volumetric light field to yield arbitrary 3D intensity distributions within the resin. This approach enables the direct formation of complex architectures, including hollow structures defined by dark interior regions, that are otherwise inaccessible with conventional illumination. Simultaneously, we design the resin’s absorption profile to balance minimal attenuation with sufficient energy deposition for structural fidelity. This enables the fabrication of millimetre-scale architectures with sub-100$\thinspace\mu$m features in a single exposure as fast as 7.5 seconds. Our approach establishes a new regime for high-speed volumetric manufacturing with throughput of 1 mm$^3$/s, with broad implications for scalable production of micro-optical components, biomedical scaffolds, and other precision-engineered mesoscale systems.
\end{abstract}

\noindent
Additive manufacturing (AM), or three-dimensional (3D) printing, has transformed the fabrication of complex volumetric structures, enabling unprecedented design freedom in fields ranging from electronics and photonics to biomedicine and microelectromechanical systems (MEMS)~\cite{jadhav2022review}. Among AM modalities, stereolithography (SLA) is distinguished by its ability to produce high-resolution features via photopolymerization of a photosensitive resin. Conventional SLA achieves 3D structuring by raster-scanning a tightly focused laser through the resin volume~\cite{somers2024physics, zyla2024frontiers}, often relying on two-photon absorption to restrict reactivity to the focal point. While this nonlinearity enables precise voxel localization, it requires high laser intensities and limits scalability~\cite{yang2014projection}.

To alleviate the necessity of high laser power and to broaden material compatibility, alternative photochemical strategies based on single-photon and two-step activation have been proposed. Single-photon SLA offers simpler optics, but is constrained to ultrathin structures due to the strong absorption and undesired out-of-plane curing~\cite{Maruo2000, do2013submicrometer}. Recently, molecular engineering has enabled two-step approaches that decouple excitation from polymerization, allowing printing with significantly lower light intensities. These include both single-color~\cite{hahn2021two} and two-color~\cite{regehly2020xolography, hahn2022light} implementations, which provide enhanced control over reaction initiation. However, across all these approaches, 3D structures are ultimately assembled through sequential exposures—either by scanning a focal point or stacking 2D projections—imposing fundamental limits on printing speed, scalability, and system complexity.

Efforts to overcome these limitations have driven the development of faster SLA variants. Digital light processing (DLP) accelerates fabrication by projecting entire layer patterns simultaneously, enabling high-throughput printing through layer-wise or parallelized exposure schemes~\cite{zhao2020engineering, ouyang2023ultrafast}. In the latter, scan speed is further improved via ultrafast lasers and optimized photoresists~\cite{ouyang2023ultrafast}, but the sequential nature of voxel or layer addition remains a bottleneck.

A related approach is tomographic 3D printing, which reconstructs entire objects volumetrically by projecting a sequence of light patterns into a rotating resin volume~\cite{loterie2020high, alvarez2025holographic, bhattacharya2021high}. By integrating light intensities over time and angle, this approach enables the rapid fabrication of centimeter-scale structures in under a minute. Further acceleration has been demonstrated via multi-axis illumination, achieving full-object formation in less than 10 seconds~\cite{shusteff2017one}. However, these methods require complex optical setups, precise calibration, and careful synchronization, posing challenges for widespread adoption.

An alternative path to continuous printing has emerged through the use of O$_2$-inhibited photopolymerization, which sustains a persistent liquid interface between the print and the exposure window~\cite{tumbleston2015continuous, janusziewicz2016layerless}. This “dead zone” eliminates the need for mechanical peeling between layers, enabling continuous growth of the printed part. Subsequent advances have extended this concept to roll-to-roll systems for high-throughput manufacturing~\cite{kronenfeld2024roll} and to microscale resolutions~\cite{hsiao2022single}. Nonetheless, these approaches still rely on repeated light exposures, which ultimately limit throughput and preclude full volumetric parallelization (Fig. 1a).

\begin{figure}[h]
\centering
\includegraphics[width=0.9\textwidth]{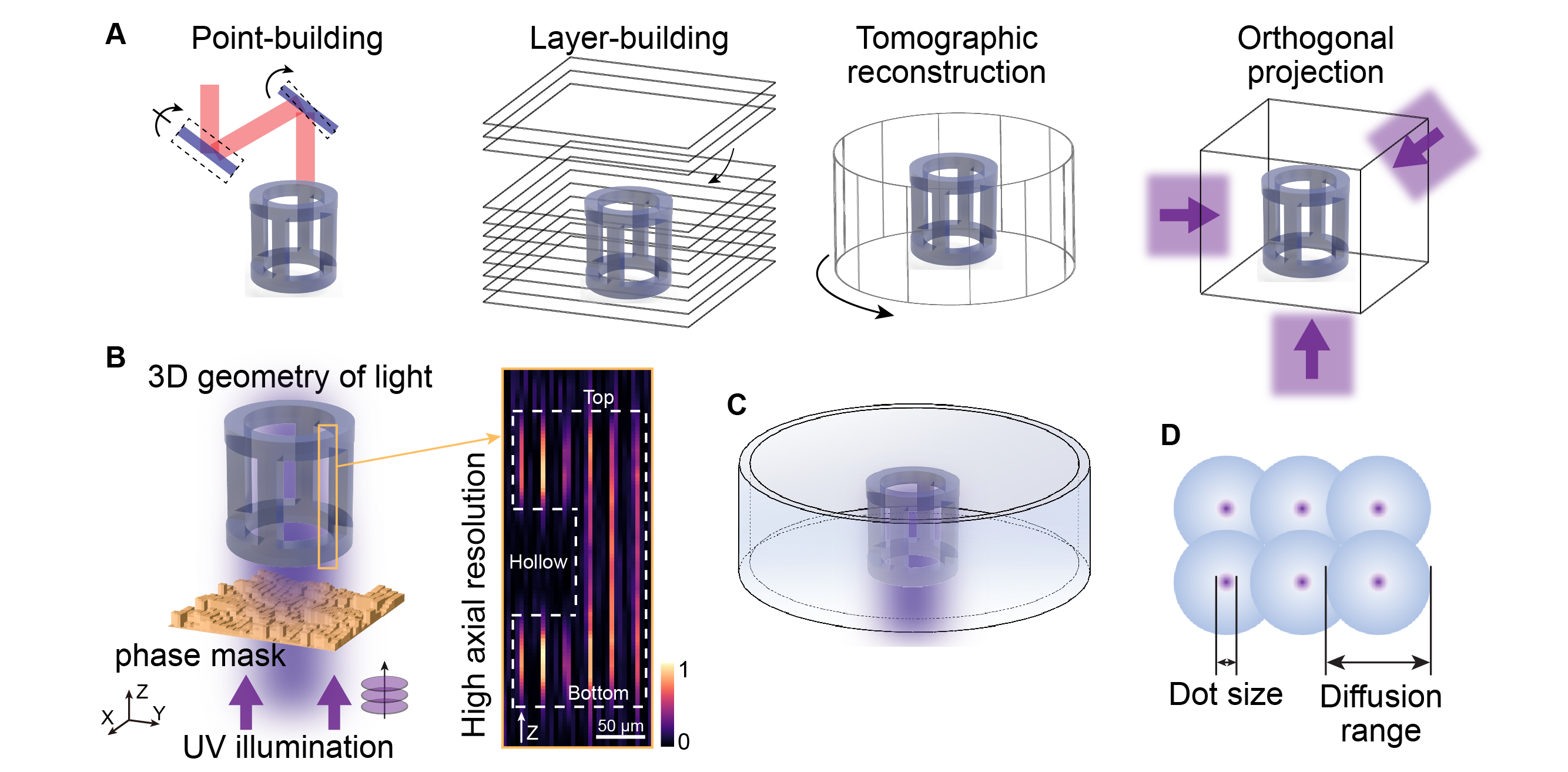}
\caption{\textbf{Single-exposure holographic additive manufacturing.}
(A) Conventional volumetric fabrication techniques rely on dynamic control elements such as galvanometric mirrors (point-scanning), stacked layer-by-layer exposure, multiple tomographic projections, or via orthogonal multi-angle beam projections. (B) In contrast, our approach generates arbitrary 3D holographic intensity distributions using a static phase mask. The mask encodes volumetric information via a spatial sampling grid for high axial resolution. In our demonstration, the mask, fabricated via grayscale lithography, measures $2.4 \times 2.4 \thinspace \text{mm}^2$ with $2\thinspace\mu$m minimum features. Right panel shows a small portion of the simulated 3D holographic light distribution of the hollow cylinder illustrating high axial resolution (see details in Fig. \ref{fig:inverse_design}). (C) The computed 3D intensity distribution is projected into a UV-curable resin, enabling simultaneous volumetric polymerization in a single exposure. (D) Mask and resin parameters are co-optimized to ensure localized light distributions (Gaussian “dots”) merge into a continuous 3D structure upon exposure.}\label{fig:intro}
\end{figure}

We introduce a volumetric additive manufacturing strategy that enables millimetre-scale 3D microstructures to be fabricated in a single holographic exposure. In contrast to conventional layer-by-layer methods, which rely on sequential light modulation and are constrained in both throughput and resolution, our approach projects the entire volumetric intensity distribution into the resin in one step. This is achieved using a single inverse-designed phase mask that encodes the desired hologram, allowing arbitrary and complex architectures to be polymerized simultaneously (Fig. ~\ref{fig:intro}).

Building on advances in 3D holographic displays, we designed and fabricated an inverse-engineered phase mask (a diffractive optical element or computer-generated hologram)~\cite{heanue1994volume, shi2021towards, gopakumar2024full} using grayscale optical lithography. When illuminated, this static phase mask sculpts the volumetric light field with high fidelity, curing the resin in a single exposure. Compared with conventional spatial light modulators, the inverse-designed mask achieves a markedly higher space–bandwidth product (SBP), enabling projection of complex volumetric patterns. As demonstration, we fabricate hollow structures ($\approx 2 \times 2 \times 2\thinspace\text{mm}^3$) with sub-100$\thinspace\mu$m features in under 8 s, underscoring both the precision and throughput of this holographic manufacturing platform.

\subsection*{Inverse design}
Inverse design refers to the process of specifying an optical element by first defining a desired target function—such as a 3D light intensity distribution—and then computationally optimizing the design to achieve it~\cite{menon2015nanophotonic, molesky2018inverse}. Prior studies have demonstrated that carefully engineered microstructures on a phase mask can be used to shape complex 3D light fields~\cite{meem2019multi, mohammad2017full, banerji2020extreme, sun2022three, lin2025diffraction}. This approach enables the generation of arbitrary volumetric intensity patterns, constrained only by diffraction and the SBP of the phase mask. Notably, since our phase masks are fabricated using lithography, they can achieve substantially higher SBP than traditional spatial-light modulators, enabling finer spatial control and larger pattern volumes.

To generate the target 3D intensity distribution, we begin by preprocessing the input geometry, as illustrated in Fig.~\ref{fig:inverse_design}. A computer-aided design (CAD) model, typically spanning $\approx 2 \times 2 \times 2 \thinspace \text{mm}^3$, is imported and axially sliced into a user-defined number of layers to ensure structural continuity. Each slice is then discretized using a Gaussian-dot-shaped sampling grid, with full width at half maximum (FWHM) 4~$\mu$m and inter-dot spacing of 24$\thinspace\mu$m (Fig.~\ref{fig:inverse_design}A). This sampling strategy enhances axial resolution and enables fine control over energy deposition. By optimizing the peak intensity and spatial arrangement of the dots, we further improve the fidelity of the reconstructed pattern. Detailed analysis including comparisons to a continuous pattern and of the influence of dot spacing, shape, and size are provided in the Supplementary Notes S1-S3. To highlight the versatility of the inverse design process in shaping complex 3D light fields, we include simulated reconstructions of standard benchmark geometries in Supplementary Note S5. 

The inverse design of the phase mask is implemented using the TensorFlow framework (v2.10, Google Inc.). In the example shown, the phase mask spans $2.4~\mathrm{mm} \times 2.4~\mathrm{mm}$, consisting of $1200 \times 1200$ pixels with $2~\mu\mathrm{m}$-wide square pixels, and a maximum height of $587~\mathrm{nm}$, optimized for a working wavelength of $405~\mathrm{nm}$. The phase modulation is defined as:
\begin{equation}
t(x, y) = \exp\left( \frac{j2\pi h(x, y)\Delta n}{\lambda} \right),
\end{equation}
\noindent where $h(x, y)$ is the height profile to be optimized, $\lambda$ is the wavelength, and $\Delta n$ is the refractive index contrast between the photoresist and air.

Light propagation from the phase mask to the resin is modeled using the angular-spectrum method across a user-defined gap (in the example shown, gap $=5~\mathrm{mm}$), yielding the intensity distributions $I(Z_i)$ at a set of axial planes. The objective of the inverse design is to minimize the discrepancy between the simulated volumetric intensity and the target 3D distribution. This is achieved by minimizing a weighted sum of mean squared errors (MSEs) across all slices:
\begin{equation}
\text{Loss} = \sum_{i=1}^{n} w(Z_i) \cdot \left(I(Z_i) - \alpha T(Z_i)\right)^2,
\end{equation}
\noindent where $T(Z_i)$ is the target intensity at depth $Z_i$, $\alpha$ is a normalization factor, and $w(Z_i)$ are slice-specific weights. The procedure for the choice of these parameters are described in Supplementary Note S7. The forward light propagation model is fully differentiable, and the loss gradient is calculated for back-propagation during optimization. The optimization is terminated when the loss function value decreases by less than 0.1\%. Optionally, to account for optical attenuation within the resin, we can apply an exponential intensity correction derived from the Beer–Lambert law (see Supplementary Note S6 for details). 

Optional intensity-stop regions may be introduced above and below the target geometry to suppress unintended polymerization outside the intended volume. The resulting axial intensity profile exhibits controlled exponential modulation within the structure and sharply attenuated boundaries at the stop zones, enabling high-fidelity volumetric fabrication with minimal overexposure. The target intensity pattern incorporated a slight bias to account for residual optical absorption in the resin, which was further reduced through optimized material formulation (described below).

\subsection*{Experiments}
\begin{figure}[h]
\centering
\includegraphics[width=0.9\textwidth]{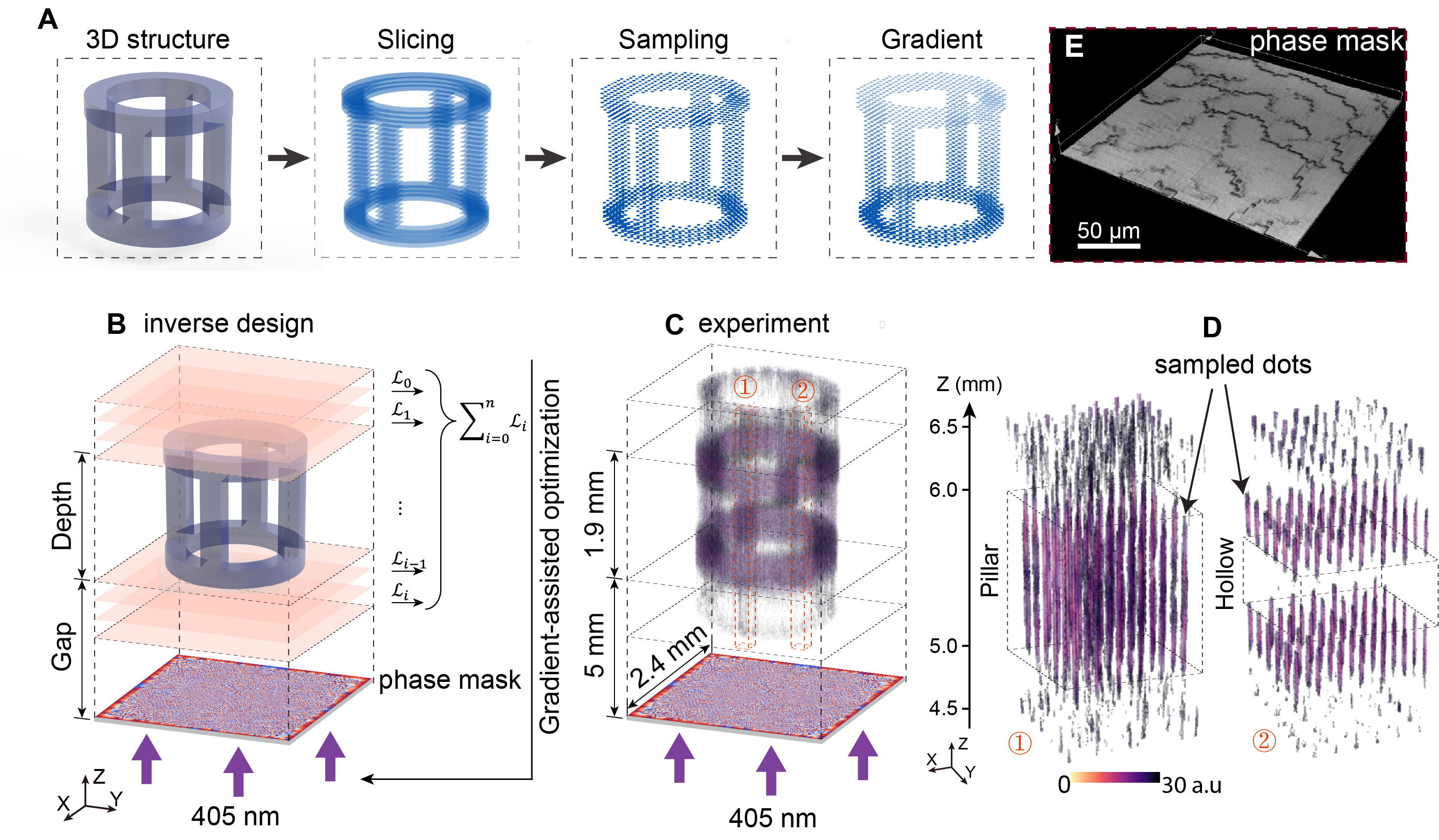}
\caption{\textbf{Inverse design of the phase mask for single-shot 3D fabrication.} (A) Workflow for generating 3D target intensities from CAD geometries.(B) Inverse design methodology to attain the target 3D intensity distribution. (C) Measured 3D holographic intensity distribution, showing solid pillars and a hollow cylindrical core. (D) Magnified intensity maps from pillar and non-pillar regions, resolving individual sampled dots and confirming the absence of light in the hollow wall. (E) Optical micrograph of the fabricated phase mask.}\label{fig:inverse_design}
\end{figure}

\begin{figure}[h]
\centering
\includegraphics[width=0.9\textwidth]{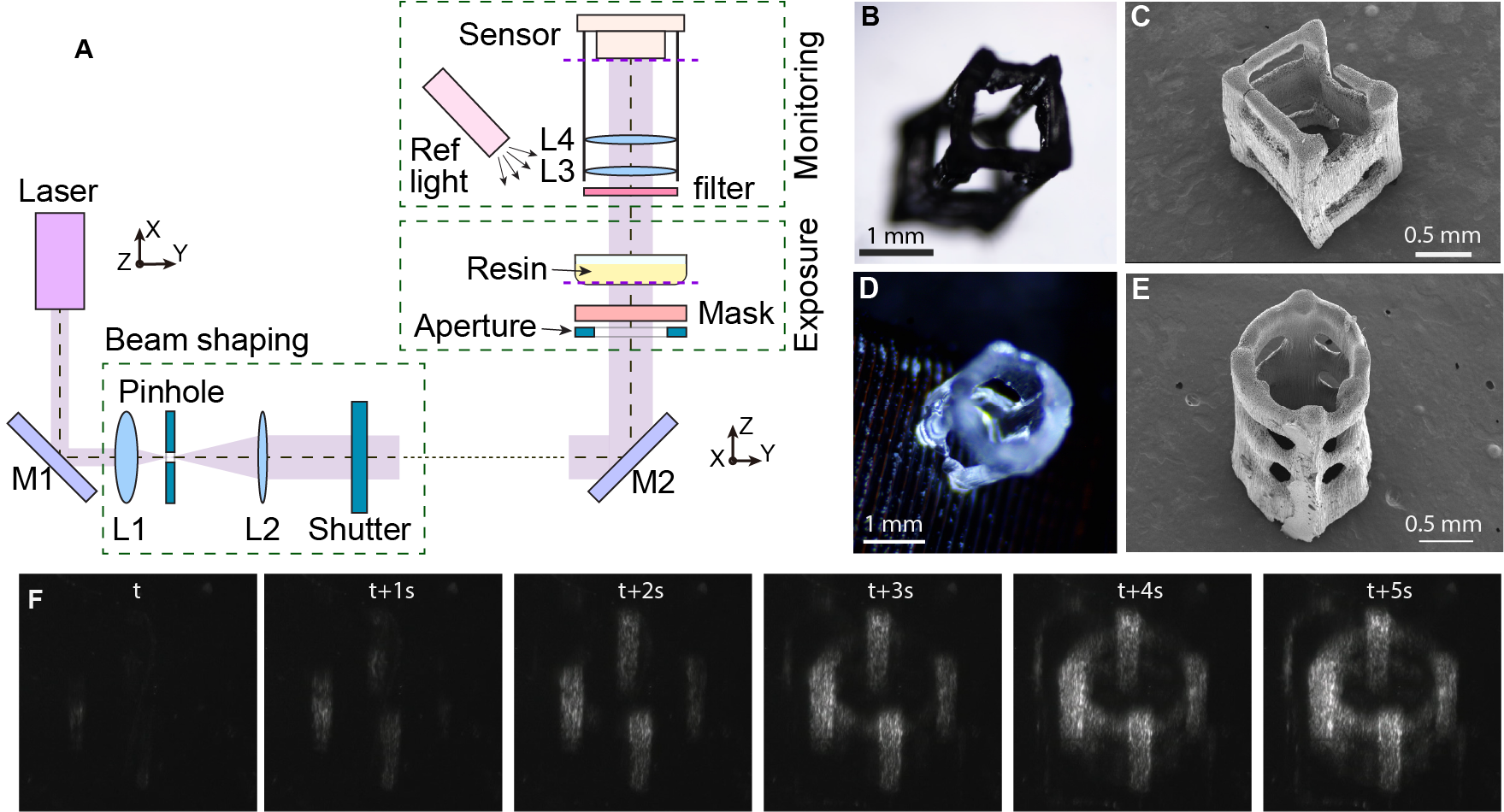}
\caption{\textbf{Single-exposure holographic 3D printing.}
A, Schematic of the experimental setup, where a shutter controls exposure duration. B–E, Optical and scanning electron micrographs of representative structures—a hollow cylinder and a cube—fabricated in a single exposure. Prints in B, C, and E used 15 s exposures, while D was produced in 7.5 s. The multiple windows visible in C and E arise when the resin thickness exceeds the designed volume. F, Time-lapse images of the hollow cylinder, recorded at 1 s intervals, reveal real-time progression of polymerization. Illumination was provided at 850 nm, with a long-pass filter before the camera to enhance contrast.}\label{fig:print}
\end{figure}

The phase mask was fabricated on a 2-inch glass wafer using grayscale lithography. Surface profiles were characterized via scanning-confocal microscopy (Olympus LEXT OLS5000). The mask was illuminated with collimated 405 nm laser light (100 mW, Edmund Optics, \#19-462) using the optical setup shown in Fig.~\ref{fig:print}. The resulting 3D light-intensity distribution was captured via axial scanning an image sensor (Supplementary Information Note S8), and is presented in Fig. ~\ref{fig:inverse_design}C. The reconstructed hologram exhibits needle-like columns aligned with the optical axis, a direct outcome of the dot-sampling strategy used in the inverse design. Magnified views of the pillar and central hollow region (Fig. ~\ref{fig:inverse_design}D) confirm spatial confinement of intensity. While some residual light appears in stop-designated regions, its magnitude is substantially lower than that in the target volume.

To avoid spurious polymerization within the hollow core of the printed structure, the optical intensity must remain below the resin’s curing threshold. We can quantify the image contrast in each layer as $K=(I_{\text{inside}}-I_{\text{outside}})/(I_{\text{inside}}+I_{\text{outside}})$, where $I_{\text{inside}}$ and $I_{\text{outside}}$ are the average intensity inside and outside the target region, respectively. From the measured 3D intensity distributions (Figs. ~\ref{fig:inverse_design}C and D), we extracted the contrast in 3 representative planes corresponding to z=5 mm (circular frame), z = 5.5 mm (pillars) and z = 6 mm (circular frame) as 21\%, 30\%, and 20\%, respectively. This contrast was sufficient for us to resolve the hollow cylinder geometry in the resin. Each pillar emerges from an array of discrete focal spots that locally overcome O$_2$ inhibition and drive photopolymerization. Crucially, background light within both the central void and peripheral regions remain below threshold, ensuring that minimal unintended resin curing occurs. The interplay between dose and exposure time can be quantitatively assessed through 2D slice imaging via a DLP-based platform (Supplementary Note S13).

High-fidelity 3D fabrication requires precise control over the volumetric distribution of light to selectively drive curing of the desired object, while suppressing polymerization in the undesired regions. In order to compensate for optical attenuation through the resin, we tuned the concentration of BAPO ($\epsilon_{405\thinspace\text{nm}}=546~\mathrm{M^{-1}cm^{-1}}$) in the monomer mixture so that the peak intensity at the incident layer matched that at approximately 2 mm in depth (Supplementary Note S6). Based on the experimentally measured greyscale of 1.1:1 for the mask, a resin was formulated with BAPO as the photoinitiator (PI) at a concentration of 0.46 mM (0.017 wt\%). Polymerization kinetics were measured via FTIR spectroscopy, where low excitation intensities result in delayed crosslinking due to increased O$_2$ inhibition. Complementary photorheology confirmed the strong dependence of curing dynamics on optical dose.

While the BAPO concentration is constrained to $0.46~\mathrm{mol/m}^3$ to limit light attenuation, higher loading can accelerate crosslinking, sharpen the polymerization threshold, and suppress O$_2$ inhibition, thereby enhancing resolution and print fidelity. Increased photoinitiator levels also improve crosslink density, yielding mechanically stiffer structures capable of supporting overhangs and complex geometries. However, elevated concentrations necessitate stronger correction for resin absorption, which can reduce image contrast. The chosen formulation therefore reflects a practical compromise between curing efficiency, resolution, and optical fidelity.

The optical setup for 3D printing is illustrated in Fig.~\ref{fig:print}A. A collimated ultraviolet (UV) laser source provides the illumination. The beam is expanded and aligned to uniformly illuminate the holographic phase mask. A square aperture is positioned upstream of the mask to block stray light and to ensure clean beam boundaries. A precision shutter is used to control the exposure time. An optical relay system projects the intensity distribution onto a CMOS sensor, enabling system alignment, pattern verification, and real-time monitoring of resin polymerization. Conjugate planes (purple dashed lines) establish 1:1 magnification between the projected light field and the sensor, permitting accurate axial mapping and precise registration of the resin volume with the 3D light distribution (see Supplementary Notes S8–S9 for alignment procedure). As shown in Fig.~\ref{fig:print}A, the phase mask is mounted at the base of the optical setup, with the resin contained in a standard petri dish positioned 5 mm above the mask plane. The optically flat, thin dish bottom minimizes distortion and preserves uniform axial resolution. Representative geometries—a hollow cylinder and a cube—are shown in Figs.~\ref{fig:print}B–E. The structure in Fig.~\ref{fig:print}D was exposed for 7.5 s, while those in Figs.~\ref{fig:print}B,C,E required 15 s due to reduced illumination intensity under a slightly modified optical configuration. All samples were developed by immersion in isopropyl alcohol at $45^\circ$C to remove uncured resin, and post-cured for 30 s using the same collimated UV source.

The exposure duration can be adaptively controlled using feedback from the monitoring system. Time-lapse images of a hollow cylinder print (Fig.~\ref{fig:print}F) illustrate the process, where regions that have polymerized appear darker owing to increased absorption relative to uncured resin (also see Supplementary Video 1). This contrast provides a direct visual marker of exposure completion. Illumination was delivered at 850 nm, with a long-pass filter placed before the CMOS sensor to enhance image contrast.

\subsection*{Discussion}
Single-exposure 3D printing has previously been demonstrated only for a narrow class of structures that coincide with optical eigenmodes in homogeneous media, such as beams carrying orbital angular momentum~\cite{yang2017direct, cheng2022high}, Bessel beams for fiber formation~\cite{cheng2019fabrication}, ring-shaped microtubules~\cite{ji2018high}, and self-focused filaments~\cite{liu2022filamented}. These approaches, while elegant, are inherently restricted to specific field symmetries and do not generalize to arbitrary volumetric geometries. By contrast, the inverse-designed holographic method introduced here enables the single-shot generation of complex 3D architectures. Our approach relies on discretely sampling a target geometry into a grid of Gaussian-like foci that rapidly converge and diverge along the propagation axis, yielding higher axial resolution than continuous intensity distributions. A systematic comparison of sampled and continuous designs is presented in Supplementary Notes~S1–S2.  

A key design challenge lies in avoiding unintended beam morphologies. For instance, sampled Gaussian spots can approximate non-diffracting Bessel beams. This effect accounts for the premature curing of vertical rods relative to horizontal features in cylinder and cube geometries, potentially exacerbated by waveguiding effects. Tilting structures relative to the optical axis mitigates these non-uniformities, as demonstrated in the tilted ``U'' design (Supplementary Fig.~S5).  

The exposure profile of this method differs fundamentally from that of conventional stereolithography. Instead of continuous intensity distributions, the phase mask produces sampled arrays of high-peak-intensity points embedded in low-intensity backgrounds. The resulting radical generation is therefore highly localized, while O$_2$ inhibition in the surrounding resin suppresses curing in low-dose regions. This natural thresholding effect enhances spatial resolution by minimizing unintended polymerization. To further suppress background curing, we incorporated TEMPO (2,2,6,6-Tetramethylpiperidine-1-oxyl), a radical scavenger, which diffuses more slowly than dissolved O$_2$, and  enables the formation of finer features (Fig.~2G).

\subsection*{Conclusion}
In summary, we have introduced a single-exposure holographic additive manufacturing platform that overcomes the intrinsic speed limitations of serial and layer-by-layer 3D printing. By combining inverse-designed phase masks with tailored resin formulations, we demonstrate the volumetric fabrication of millimeter-scale architectures with sub-100 $\mu$m features in a single exposure of less than 8 s. This approach achieves a printing throughput of approximately 1 mm$^3$/s—one to two orders of magnitude faster than leading tomographic or projection-based methods—while maintaining micron-scale feature resolution. Beyond establishing a new performance regime in additive manufacturing, the method is inherently scalable: phase masks fabricated over larger areas and sub-wavelength diffractive-optical elements promise further increases in space–bandwidth product, axial resolution, and design versatility. These advances open a path toward high-speed, parallelized production of functional mesoscale systems, ranging from optical and photonic components to biomedical scaffolds and architected materials.

\newpage

\clearpage 
\bibliography{science_template} 
\bibliographystyle{sciencemag}

%
%
%
%
%
%

\newpage
\section*{Acknowledgments}
The authors acknowledge support from Joseph Jacob and Brian Baker for fabrication of the holograms. 
\paragraph*{Funding:}
All authors were funded by NSF Future Manufacturing grant \#2229036.
\paragraph*{Author contributions:}
Inverse design, mask fabrication, experiments: DL. 
Experiments, resin formulation and characterization: XC, COD, J-W. K., K. S. M. 
Experiments: KSL. 
Mask characterization, Experiments: AM. 
Conceptualization: C-H.C, MC, ZAP, RM. 
All authors edited the manuscript. 

\paragraph*{Competing interests:}
There are no competing interests to declare.
\paragraph*{Data and materials availability:}
All data needed to evaluate the conclusions in the paper are present in the paper and/or the Supplementary Materials.


\newpage


\renewcommand{\thefigure}{S\arabic{figure}}
\renewcommand{\thetable}{S\arabic{table}}
\renewcommand{\theequation}{S\arabic{equation}}
\renewcommand{\thepage}{S\arabic{page}}
\setcounter{figure}{0}
\setcounter{table}{0}
\setcounter{equation}{0}
\setcounter{page}{1} 

\subsection*{Materials and Methods}

\textbf{Fabrication of phase masks:} Phase masks were fabricated via grayscale optical lithography. A 2-inch soda-lime glass substrate was spin-coated with a positive-tone photoresist (MICROPOSIT S1813 G2) at 2000 rpm for 60 s, followed by a soft bake at 110 $^0$C for 2 minutes. Precise control of the resulting topography requires prior calibration of the exposure dose–height relationship, enabling accurate translation of grayscale patterns into continuous surface relief structures. Grayscale optical lithography was performed using a direct-write laser lithography system (DWL 66+, Heidelberg Instruments), which modulates the local exposure dose according to the inverse-designed phase profile. Following exposure, the substrate was post-baked at 50 $^0$C for 1 minute and developed in a 1:1 dilution of AZ developer for 1 minute, yielding the final surface-relief structure.

\noindent\textbf{Resin preparation:} All reagents were of analytical grade and used as received unless otherwise specified. Phenylbis(2,4,6-trimethylbenzoyl)phosphine oxide (BAPO, 99\%, AmBeed) was recrystallized prior to use. 2-Phenoxyethyl acrylate (PEA) was provided by Osaka Organic Chemical Industry Ltd., and dipentaerythritol penta-/hexa-acrylate ($\leq$650 ppm MEHQ inhibitor, Sigma Aldrich) served as the crosslinker. The resin formulation consisted of dipentaerythritol penta-/hexa-acrylate (90 wt\%), PEA (10 wt\%), and BAPO, mixed in a glass vial on a heat shaker at 50 $^\circ$C for 1 h. To ensure homogeneity and remove bubbles, the mixture was centrifuged in a speed mixer at 1600~rpm for 10~min. The BAPO concentration was adjusted according to the grayscale holographic pattern, compensating for optical attenuation using the Beer–Lambert law. 

\end{document}